\documentclass{article}
\usepackage{cite}
\usepackage{spconf,amsmath,graphicx,hyperref}
\usepackage{amsmath,amssymb,amsfonts}
\usepackage{algorithmic}
\usepackage{graphicx}
\usepackage{textcomp}
\usepackage{xcolor}
\usepackage{hyperref}
\usepackage{enumitem}
\usepackage{xurl}
\usepackage{booktabs}
\usepackage{multirow}
\def\BibTeX{{\rm B\kern-.05em{\sc i\kern-.025em b}\kern-.08em
    T\kern-.1667em\lower.7ex\hbox{E}\kern-.125emX}}
\usepackage{xcolor}
\definecolor{orange}{RGB}{255,107,0}

\title{Discover Fast Power Allocation Solution for Multi-Target Tracking via AlphaEvolve Evolution
}

\name{{
    Zhenkang Hou\textsuperscript{\dag}, \quad
    Wenqiang Pu\textsuperscript{\ddag}, \quad
    Junkun Yan\textsuperscript{\dag}, \quad
    Rui Zhou\textsuperscript{\ddag}, \quad
    Hongwei Liu\textsuperscript{\dag}, \quad
}}

\address{\textsuperscript{\dag}National Laboratory of Radar Signal 
Processing, Xidian University, Xian, China\\
  \textsuperscript{\ddag} Shenzhen Research Institute of Big Data, The Chinese University of Hong Kong, Shenzhen, China
  }

\begin{document}
\ninept
\maketitle

\begin{abstract}
Efficient radar resource allocation is a fundamental yet computationally challenging problem, as optimal solutions typically require iterative optimization with high complexity. Motivated by the need for real-time scheduling, robust generalization, and low data dependency, this paper proposes a novel paradigm that leverages large language model (LLM)-guided evolutionary search (AlphaEvolve) to autonomously discover a closed-form power allocation solution for multi-target tracking. The approach encodes high-dimensional radar states into physically inspired features, then evolves a compact and interpretable scoring function, which is transformed to feasible power allocations via a deterministic constraint-satisfying transformation. Extensive experiments demonstrate that the discovered closed-form solution achieves near-optimal tracking accuracy (average relative performance loss of only $1.51\%$), reliable generalization across diverse scenarios and target counts, and over three orders of magnitude speedup compared to conventional iterative solvers. These results highlight the potential of LLM-guided symbolic search to revolutionize not only radar resource management but also broader classes of engineering optimization problems.

% This paper proposes a closed-form power allocation  for single-radar multi-target tracking based on the AlphaEvolve framework. Specifically, we extract physically inspired features from high-dimensional radar state matrices, evolve a closed-form power scoring function with these physically inspired features as independent variables through the AlphaEvolve framework, and transform the resulting scores into a feasible power allocation strategy via a deterministic constraint-satisfying power transformation. Experimental results demonstrate that the discovered formula achieves near-optimal accuracy, strong generalization capability and high computational efficiency, and exhibits stable performance without filter divergence in long-term dynamic closed-loop tracking.
\end{abstract}

\begin{keywords}
radar resource allocation, AlphaEvolve, large language model, convex optimization
\end{keywords}

\section{Introduction}
Radar resource management is critical for ensuring effective operation of radar systems in complex and dynamic environments~\cite{b1,b2}. It involves making real-time decisions on how to allocate limited resources, such as transmit power, revisit time, and beam directions, among multiple targets to optimize overall sensing performance~\cite{b2,b4,b5,b6,b7}.

In the literature, many existing works formulate the radar resource allocation problem within an optimization framework~\cite{b3,b4,b5,b6,b7}. Typical design objectives include maximizing tracking accuracy~\cite{b8}, minimizing total transmit power consumption, and optimizing low probability of intercept performance. These problems are subject to practical physical constraints, such as the upper and lower bounds of transmit power, dwell time, and signal bandwidth, as well as the minimum signal-to-noise ratio (SNR) requirement for reliable target detection~\cite{b4,b5,b6,b7}. The resulting problems are usually nonlinear, nonconvex, or mixed-integer programming problems, though in some cases they can be transformed into convex ones~\cite{b7}. To solve these optimization problems, various iterative algorithms have been proposed, including gradient projection~\cite{b4}, particle swarm optimization~\cite{b5}, sequential quadratic programming~\cite{b6}, dual ascent, and block coordinate descent methods~\cite{b7}. While these methods can achieve optimal or suboptimal solutions, they typically require a large number of iterations to converge.

Since an approximate solution is generally acceptable in engineering practice, data-driven approaches leveraging deep neural networks (DNNs) have been explored to directly learn the transformation from radar states to resource allocations, shifting the computational burden to an offline training phase to achieve low-latency online inference~\cite{b9,b10,b11,b12,b13}. However, the inherent black-box nature of DNNs limits the interpretability of their decisions~\cite{b12,b13}, and their generalization degrades under distribution shift when operational conditions deviate from the training scenarios~\cite{b12,b13}.

Another different paradigm is offered by symbolic regression~\cite{b14,b15}, which searches for compact algebraic expressions that explicitly describe the relationships in data. Unlike neural-network fitting, symbolic regression yields human-readable formulas with clear physical interpretation, making it an attractive tool for discovering analytical approximations to engineering optimization problems. The integration of large language models with symbolic regression has further extended the reach of expression discovery to complex engineering domains~\cite{b16}. AlphaEvolve~\cite{b17} proposed by Google DeepMind is a landmark work that has significantly advanced the field of evolutionary program search. As a substantial enhancement over prior systems like FunSearch, AlphaEvolve orchestrates an autonomous closed-loop pipeline. It has successfully discovered algorithms that surpass decades of known results in matrix multiplication, combinatorial optimization, and hardware circuit design~\cite{b17}. However, its potential in solving real-time radar resource allocation problems remains unexplored to date.

In this paper, we focus on the single-radar multi-target tracking power allocation problem, where the optimal solution relies on computationally expensive matrix inversion operations, and no closed-form expression can be derived via conventional convex optimization analysis. Inspired by the advantages of symbolic regression, we reformulate this resource allocation problem into a physics-driven symbolic search task. Unlike the conventional data-fitting paradigm of symbolic regression~\cite{b14,b15,b16}, which focuses on fitting predefined input-output labels, we embed the physical priors of radar tracking into the AlphaEvolve framework. This transforms AlphaEvolve from a pure data-fitting tool into a physics-guided symbolic optimizer. Thus, our paradigm fundamentally bypasses the massive data dependency typical of deep learning approaches. Consequently, it avoids performance degradation under distribution shifts and ensures robust generalization. By efficiently navigating the vast mathematical expression space, AlphaEvolve autonomously discovers a white-box, closed-form power scoring function that achieves near-optimal accuracy while reducing the online computational complexity to $\mathcal{O}(N)$.

\section{Radar Power Allocation Formulation}

Modern radar systems often need to track multiple targets in real time, requiring efficient allocation of limited transmit power among all targets to maximize overall tracking performance. Consider a single-radar system tracking $N$ targets in a 2D Cartesian coordinate system. At each scheduling instant, the radar allocates its total transmit power $P$ among the $N$ targets to obtain range and azimuth measurements for each one. The state of the $i$-th target is described by $\mathbf{x}_i = (x_i, \dot{x}_i, y_i, \dot{y}_i)$, where $(x_i, y_i)$ and $(\dot{x}_i, \dot{y}_i)$ are the position and velocity components respectively.

The radar receives noisy measurements in terms of range and azimuth for each target, whose accuracy depends on the allocated power. The measurement errors in range and azimuth are assumed to be statistically independent. Therefore, the measurement covariance for the $i$-th target can be modeled as a diagonal matrix, given by:
\begin{equation}
\boldsymbol{\Sigma}^{i} = \mathrm{diag}(\sigma_{r,i}^2,\, \sigma_{\theta,i}^2)
\label{eq:Sigma}
\end{equation}
where $\sigma_{r,i}^2$ and $\sigma_{\theta,i}^2$ are the variances of range and azimuth measurements, respectively. According to the radar equation~\cite{b18}, these variances are inversely proportional to the received signal-to-noise ratio (SNR), which is a linear function of the allocated power $p_i$ given as
\begin{equation}
\sigma_{r,i}^2 \propto {p_i^{-1} \gamma_{r,i}}, \qquad \sigma_{\theta,i}^2 \propto {p_i^{-1} \gamma_{\theta,i}}
\label{eq:variance}
\end{equation}
where $\gamma_{r,i}$ and $\gamma_{\theta,i}$ encapsulate the effects of target range, radar cross section (RCS), antenna gain, and other system parameters. Intuitively, increasing $p_i$ directly improves the measurement accuracy for the $i$-th target. To quantify the achievable tracking accuracy, the BCRLB is commonly adopted in the literature as the performance metric. For the $i$-th target, the Bayesian information matrix is
\begin{equation}
\mathbf{J}_i(p_i) = \mathbf{J}_{\text{prior},i} + p_i \cdot \widetilde{\mathbf{J}}_{d,i}
\label{eq:Ji}
\end{equation}
where $\mathbf{J}_{\text{prior},i}$ is the prior information matrix (i.e., the inverse of the predicted state covariance matrix) obtained from the tracking filter prediction step (e.g., using an Extended Kalman Filter (EKF)~\cite{b19}, and $\widetilde{\mathbf{J}}_{d,i}$ is the unit-power normalized measurement information matrix. The explicit formulations of the prior information matrix $\mathbf{J}_{\text{prior},i}$ and the normalized measurement information matrix $\widetilde{\mathbf{J}}_{d,i}$ are derived in detail in Appendix~\ref{app:BIM}. The BCRLB for the $i$-th target is
\begin{equation}
\mathrm{BCRLB}_i(p_i) = \sqrt{\mathrm{Tr}_p\!\left(\mathbf{J}_i^{-1}(p_i)\right)}
\label{eq:bcrlb}
\end{equation}
where $\mathrm{Tr}_p(\cdot)$ denotes the trace over the position components of the posterior covariance matrix. A smaller $\mathrm{BCRLB}_i$ indicates higher potential tracking accuracy. The power allocation problem is then formulated as minimizing the total weighted position BCRLB across all targets:
\begin{equation}
\begin{aligned}
\mathcal{P}:\quad & \min_{\{p_i\}} \sum_{i=1}^{N} w_i \cdot \mathrm{BCRLB}_i(p_i)=F(\mathbf{p}) \\
& \text{s.t.}\quad \sum_{i=1}^{N} p_i = P,\quad p_i \geq p_{\rm{min}},\; \forall\, i
\end{aligned}
\label{eq:opt}
\end{equation}
where $w_i$ is the priority weight for the $i$-th target, and $p_{\rm{min}}$ is the minimum power to guarantee basic illumination. This objective seeks the best overall tracking precision under limited resources, with higher-weight targets contributing more to the cost.

Problem $\mathcal{P}$ is convex and admits a unique global optimum. However, the optimal solution cannot be written in closed form due to the nontrivial dependence of $p_i$ inside the matrix inversion. Standard numerical solvers such as the interior-point method (IPM) rely on an iterative procedure to reach convergence. As the number of targets $N$ grows, the total execution time accumulated over these iterations increases substantially. This limitation motivates us to investigate whether an iteration-free, efficient resource allocation algorithm can be developed to closely approximate the optimum with minimal computational complexity.

\section{Power Allocation Strategy Discovery}

This section presents the proposed paradigm to discover a closed-form power allocation solution. First, we extract physically inspired features from the high-dimensional radar state matrices, to construct a compact feature vector for each target in Section~\ref{feature_extraction}. Then, we leverage the AlphaEvolve framework to evolve a closed-form scoring function that maps these features to power priority scores in Section~\ref{alphaevolve_search}, and the closed-form scoring function is presented in detail in Section~\ref{scoring_function}. Finally, in Section~\ref{deterministic_transformation}, we apply a deterministic constraint-satisfying transformation to convert the raw scores into feasible power allocations that strictly adhere to all physical constraints.

\subsection{Physically Inspired Feature Extraction}
\label{feature_extraction}
\begin{table*}[htbp]
    \centering
    \caption{Definitions of the Feature Vector}
    \label{tab:features}
    \renewcommand{\arraystretch}{1.3}
    \begin{minipage}{14cm}
    \begin{tabular}{lllp{6cm}}
        \toprule
        \textbf{Category} & \textbf{Feature} & \textbf{Description} & \textbf{Mathematical Definition} \\
        \midrule
        \multirow{8}{*}{\shortstack{Target-Level \\ Physical \& \\ Observation}} 
        & $X_1$ & Total number of targets & $N$ \\
        & $X_2$ & Target range & $R_i$ \\
        & $X_3$ & Priority weight & $w_i$ \\
        & $X_4$ & RCS & $\sigma_i$ \\
        & $X_5$ & Full-power SNR & $P G^2 \lambda^2 \sigma_i / ((4\pi)^3 R_i^4 F k T_0 B_n)$ \\
        & $X_6$ & Prior position variance & $\text{Tr}_p(\mathbf{J}_{\text{prior},i}^{-1})$ \\
        & $X_7$ & Full-power measurement info. & $P \text{Tr}_p(\widetilde{\mathbf{J}}_{d,i})$ \\
        & $X_8$ & Information ratio & $\text{Tr}_p(\mathbf{J}_{\text{prior},i}) / \text{Tr}_p(\widetilde{\mathbf{J}}_{d,i})$ \\
        \midrule
        \multirow{2}{*}{\shortstack{Demand \\  Factor}} 
        & $X_9$ & Absolute demand factor & Eq.~\eqref{eq:Di}\\
        & $X_{10} \sim X_{12}$ & Global statistical features of $X_9$ & Mean, Max, and Std. Dev. of $\{X_{9,j}\}_{j=1}^N$ \\
        \midrule
        \multirow{4}{*}{\shortstack{Marginal \\ Benefit}} 
        & $X_{13}$ & Baseline marginal benefit & Eq.~\eqref{eq:MR_base} \\
        & $X_{14} \sim X_{16}$ & Global statistical features of $X_{13}$ & Mean, Max, and Std. Dev. of $\{X_{13,j}\}_{j=1}^N$ \\ 
        & $X_{17}$ & Cliff marginal benefit & Eq.~\eqref{eq:MR_cliff} \\ 
        & $X_{18} \sim X_{20}$ & Global statistical features of $X_{17}$ & Mean, Max, and Std. Dev. of $\{X_{17,j}\}_{j=1}^N$ \\
        \bottomrule
    \end{tabular}
    \vspace{4pt} 
    \footnotesize 
    \textit{Note:} The definitions of the terms $P, G, \lambda, \sigma_i, R_i, F, k, T_0,$ and $B_n$
     are detailed in Table~\ref{tab:radar_parameters}.
    \end{minipage}
\end{table*}

Since the state information is encoded in the posterior information matrix $\mathbf{J}_i(p_i)$, directly using its elements as raw features would obscure key domain knowledge. Instead, as mathematically proved by the Karush-Kuhn-Tucker (KKT) conditions in Appendix~\ref{app:kkt_features}, the optimal power allocation for a target is highly dependent on its baseline tracking error and its measurement gradient. Therefore, we focus on extracting a compact feature vector $\mathbf{u}_i = [X_1, X_2, \dots, X_{K}]^T$ for each target that preserves this physical insight. The complete definitions and mathematical formulations of these $K=20$ features are explicitly summarized in Table~\ref{tab:features}, and the core feature components are constructed as follows.
\begin{itemize}[leftmargin=0pt]
    \item \textbf{Absolute demand factor} $D_i$: This scalar quantifies the inherent tracking difficulty of the $i$-th target under equal resource sharing at the operating point $\bar{p} = P/N$:
    \begin{equation}
    D_i = w_i \cdot \sqrt{\mathrm{Tr}_p\!\left(\mathbf{J}_i^{-1}(\bar{p})\right)}
    \label{eq:Di}
    \end{equation}
    A larger $D_i$ reflects a higher priority, unfavorable observation geometry or large prior uncertainty indicates a greater demand for power.

    \item \textbf{Baseline and cliff marginal benefit} $M_{\text{base},i}$ and $M_{\text{cliff},i}$: These scalars quantify the rate of reduction in tracking error per unit of additional power at two critical allocation boundaries. Specifically, the former captures the marginal efficiency of power investment at the equal-allocation allocation $\bar{p}$, while the latter captures the marginal efficiency of power investment at the minimum power limit:
    \begin{equation}
    M_{\text{base},i} = \frac{w_i \cdot \mathrm{Tr}_p\!\left(\mathbf{J}_i^{-1}(\bar{p}) \cdot \widetilde{\mathbf{J}}_{d,i} \cdot \mathbf{J}_i^{-1}(\bar{p})\right)}{2\,\sqrt{\mathrm{Tr}_p\!\left(\mathbf{J}_i^{-1}(\bar{p})\right)}}
    \label{eq:MR_base}
    \end{equation}
    
    \begin{equation}
    M_{\text{cliff},i} = \frac{w_i \cdot \mathrm{Tr}_p\!\left(\mathbf{J}_i^{-1}(p_{\text{min}}) \cdot \widetilde{\mathbf{J}}_{d,i} \cdot \mathbf{J}_i^{-1}(p_{\text{min}})\right)}{2\,\sqrt{\mathrm{Tr}_p\!\left(\mathbf{J}_i^{-1}(p_{\text{min}})\right)}}
    \label{eq:MR_cliff}
    \end{equation}
    Mathematically, $M_{\text{base},i}$ and $M_{\text{cliff},i}$ are precisely the first-order partial derivatives of the $i$-th target's weighted BCRLB with respect to $p_i$ at $\bar{p}$ and  $p_{\text{min}}$, respectively. Through the matrices $\mathbf{J}_{\text{prior},i}$ and $\widetilde{\mathbf{J}}_{d,i}$, they implicitly encode the target's priori information into two compact scalars.
    \item \textbf{Global statistical features} ($X_{10} \sim X_{12}$, $X_{14} \sim X_{16}$, $X_{18} \sim X_{20}$): To provide scenario-level context and awareness of global resource competition, the aggregate statistics (mean, maximum, and standard deviation) are computed across all targets for the aforementioned demand factor and marginal benefit features.
\end{itemize}

\subsection{AlphaEvolve Search Scheme for Radar Power Allocation}
\label{alphaevolve_search}
\begin{figure}[htbp]
    \centering
\includegraphics[width=0.5\textwidth]{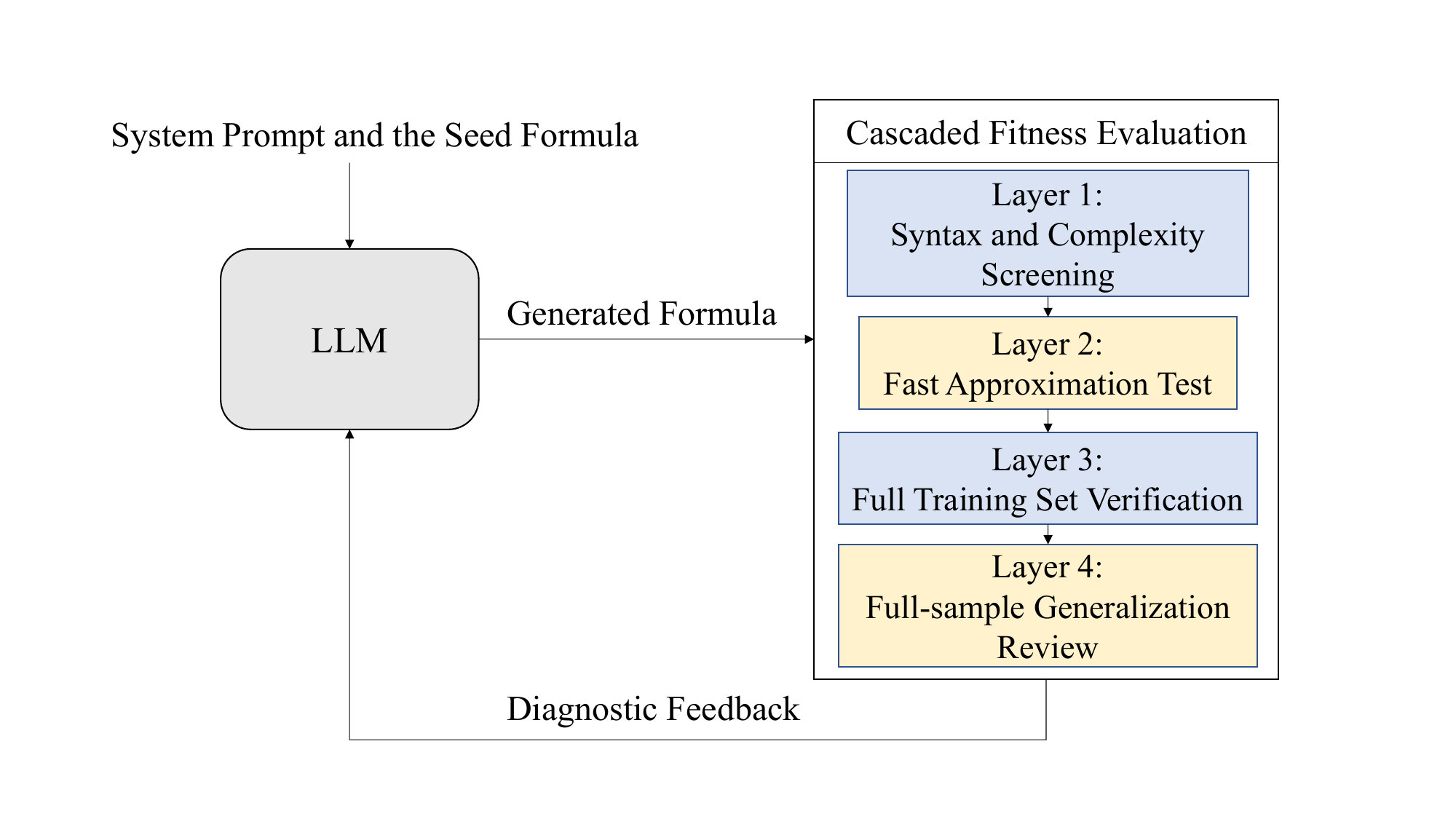}
    \caption{AlphaEvolve framework for radar power allocation.}
    \label{fig:alphaevolve_framework}
\end{figure}

The remaining issue is how to discover a closed-form scoring function $S_i = f(\mathbf{u}_i)$ that closely approximates the optimal solution across diverse scenarios with $\mathcal{O}(N)$ complexity. This is where the AlphaEvolve framework comes into play, guiding an evolutionary search over the space of algebraic expressions to find a compact formula that balances accuracy, generalization, and simplicity. As illustrated in Fig.~\ref{fig:alphaevolve_framework}, AlphaEvolve operates as a closed-loop evolutionary system with three components, initialization, cascaded fitness evaluation, and diagnostic feedback.

\textbf{Initialization.}
The system prompt instructs the LLM to act as a domain expert in radar resource management, defines the search objective, which is to discover a closed-form scoring function that approximates the optimal solution, provides a complete mathematical description of the task including the feature vector $\mathbf{u}_i$, and restricts the operator library to 12 hardware-friendly operators: $\{+,\,-,\,{\times},\,/,\,\sqrt{\cdot},\,\ln,\,\exp,\,\max,\,\min,\,\mathrm{pow},\,\tanh,\,|\cdot|\}$.
The evolutionary search is seeded with a physically motivated formula:
\begin{equation}
S_i^{(0)} = \left(w_i \cdot \sqrt{\mathrm{Tr}_p\!\left(\mathbf{J}_i^{-1}(\bar{p})\right)}\right)^{2/3} = D_i^{2/3}
\label{eq:seed}
\end{equation}
which scores each target by its tracking demand under equal power sharing, providing a meaningful starting point for the search.

\textbf{Cascaded fitness evaluation.}
Evaluating every candidate formula on the full dataset is computationally prohibitive. We therefore adopt a four-layer cascaded evaluation strategy that progressively filters candidates with increasing evaluation cost.

\textit{Layer~1 (Syntax and complexity screening).} Each candidate formula is parsed into an abstract syntax tree (AST). Candidates containing syntax errors or whose AST node count exceeds a predefined threshold are rejected immediately, before any numerical evaluation.

\textit{Layer~2 (Fast approximation test).} Surviving candidates are evaluated on a small validation subset ($30$ problems instances). The optimal solution $\mathbf{p}^*=(p_1^*,p_2^*,\ldots,p_N^*)$ serves as the reference optimum. The relative performance loss is defined as:
\begin{equation}
\Delta L_{\%} = \frac{F(\mathbf{p})}{F(\mathbf{p}^*)} \times 100\%
\label{eq:delta_L}
\end{equation}
Candidates whose $\Delta L_{\%}$ exceeds $300\%$ are rejected at this stage.

\textit{Layer~3 (Full training set verification).} Candidates that survive the fast approximation test are evaluated to compute the mean training loss $\mathcal{L}_1$ ($N = 10$--$30$) which is the mean $\Delta L_{\%}$ on training scenarios ($N = 10$--$30$). If $\mathcal{L}_1$ exceeds $250\%$, the evaluation is terminated early. To save computational resources, the candidate skips the final evaluation layer and is directly assigned a heavy predetermined penalty for the mean generalization loss ($\mathcal{L}_2$).

\textit{Layer~4 (Full-sample generalization review).} Candidates with satisfactory training performance (passing Layer~3) are assessed on the held-out large-scale scenarios ($N = 80$--$100$) to compute the true mean generalization loss $\mathcal{L}_2$, thereby enforcing cross-scale generalization.

For any candidate that passes Layer~2 (whether terminated early in Layer~3 or completing Layer~4), the overall multi-objective loss is formulated as:
\begin{equation}
\mathcal{L} = \alpha\,\mathcal{L}_1 + \beta\,\mathcal{L}_2 + \gamma\,\mathrm{Complexity}(\mathrm{AST})
\label{eq:loss}
\end{equation}
where $\mathrm{Complexity}(\mathrm{AST})$ penalizes formula length to promote parsimony, and scalar weights $\alpha$, $\beta$, $\gamma > 0$ balance the three objectives. The final fitness score returned to the AlphaEvolve framework is defined as $F = \frac{1}{1 + \mathcal{L}}$, with $F \to 1$ indicating near-optimal, compact, and generalizable performance.

\textbf{Diagnostic feedback.}
After each evaluation, structured feedback is returned to the LLM, including $\mathcal{L}_1$, $\mathcal{L}_2$, $\mathrm{Complexity}(\mathrm{AST})$, $F$, and any runtime errors. This feedback guides the LLM to propose targeted structural modifications in the next generation, steering the search toward physically consistent and computationally efficient expressions.

\subsection{Closed-Form Scoring Function}
\label{scoring_function}
After extensive evolutionary search over thousands of candidate expressions, the following scoring function is identified as the optimal trade-off between approximation accuracy and analytical simplicity 
\begin{equation}
S_i = \max\!\left(\left(\eta_i\,\zeta_i\right)^{0.495},\; 10^{-6}\right)
\label{eq:Si}
\end{equation}
where $\eta_i = {D_i}/{\bar{D}}, \ 
\zeta_i = M_{\text{base},i} / {\overline{M_{\text{base}}}}$. Here, $\bar{D}$ and $\overline{M}_{\text{base}}$ explicitly denote the means of the absolute demand factor and baseline marginal benefit across all $N$ targets in the scenario, respectively. This expression~\eqref{eq:Si} encodes three design principles. First, dividing by $\bar{D}$ and $\overline{M_{\text{base}}}$ ensures each score reflects a target's relative standing within the current scenario rather than its absolute magnitude, yielding robustness across scenarios of varying target count and difficulty. Second, the exponent $0.495\approx 1/2$ compresses the score dynamic range, preventing high-demand targets from monopolizing power while preserving nontrivial allocation for low-priority ones. Finally, the $\max(\cdot,10^{-6})$ in~\eqref{eq:Si} guarantees a strictly positive score for every target. 

It is worth noting that while the extracted feature space contains 20 dimensions (as defined in Table~\ref{tab:radar_parameters}), the evolutionary search ultimately converged to the closed-form scoring function ~\eqref{eq:Si} relying solely on the demand factor and the marginal benefit. This feature selection mechanism implies that these two categories of features capture the physical information required for optimal power allocation.

During the AlphaEvolve search process, we indeed discovered suboptimal formulas incorporating additional features. For instance, a high-fitness candidate formula integrating the cliff marginal benefit ($M_{\text{cliff},i}$) and the information ratio ($X_{8,i}$) was formulated as:
\begin{equation}
S_i^{\text{sub}} = \max\!\left(\left( \mu_i \, \beta_i \, A_i \right)^{0.493},\; 10^{-6}\right)
\label{eq:Si_sub}
\end{equation}
where $\mu_i = {D_i}/{\overline{D}}$, $\beta_i = {M_{\text{base},i}}/{\overline{M_{\text{base}}}} + 0.35 \, {M_{\text{cliff},i}}/{\overline{M_{\text{cliff}}}}$, and $A_i = 1 + 0.082 \tanh\left({X_{8,i}}/{\overline{D}}\right)$. Here, $\overline{M}_{\text{cliff}}$ denotes the means of the cliff marginal benefit across all $N$ targets in the scenario.

Cascaded fitness evaluation revealed that incorporating these additional features in \eqref{eq:Si_sub} did not yield any performance gain. Instead, \eqref{eq:Si_sub} exhibited a higher mean training loss $\mathcal{L}_1$ and mean generalization loss $\mathcal{L}_2$ compared to \eqref{eq:Si}, while having higher $\mathrm{Complexity}(\mathrm{AST})$. This indicates that forcefully integrating secondary features introduces structural redundancy rather than useful information, ultimately degrading the approximation capability. Consequently, \eqref{eq:Si} was selected as the closed-form scoring function.

% {\red [why other features are not included in the final formula, we should expalin this, a btter way is show other formualtes contains other features, but they are more complex and have worse performance.]}

\subsection{Deterministic Constraint-Satisfying Power Transformation}
\label{deterministic_transformation}
Given the scores $\{S_i\}$, the final power allocation strategy $\{p_i\}$ must satisfy the feasibility constraints $p_i \geq p_{\min}$ and $\sum_{i=1}^{N} p_i = P$. Let $\mathcal{K} = \bigl\{i : P\cdot S_i / {\textstyle\sum_k S_k} \leq p_{\min}\bigr\}$ denote the set of targets whose proportional scores fall below the minimum power threshold, and let $P_{\mathrm{res}} = P - |\mathcal{K}|\,p_{\min}$ be the power remaining after assigning $p_{\min}$ to all clamped targets. The feasible allocation is then given by:
\begin{equation}
p_i = \begin{cases}
p_{\min}, & i \in \mathcal{K} \\[4pt]
\displaystyle P_{\mathrm{res}}\cdot {S_i}/{\displaystyle\sum_{j \notin \mathcal{K}} S_j}, & i \notin \mathcal{K}
\end{cases}
\label{eq:transformation}
\end{equation}
By construction, this transformation satisfies both the equality constraint $\sum_i p_i = P$ and the inequality constraints $p_i \geq p_{\min}$ for all $i$, and requires no iterative procedure. The entire pipeline, feature extraction~\eqref{eq:Di}--\eqref{eq:MR_cliff}, scoring~\eqref{eq:Si}, and constrained transformation~\eqref{eq:transformation}, consists exclusively of vectorized algebraic operations, yielding an end-to-end computational complexity of $\mathcal{O}(N)$, in contrast to the $\mathcal{O}(N^3)$ per-iteration cost of the IPM.

\section{Verification and Performance Evaluation}

This section constructs a multi-target tracking simulation environment to verify the effectiveness of the closed-form scoring function \eqref{eq:Si}. Evaluation is conducted from four aspects: static approximation accuracy, large-scale scenario generalization, computational efficiency, and dynamic closed-loop tracking. 

\begin{figure}[ht]
    \centering
\includegraphics[width=0.45\textwidth]{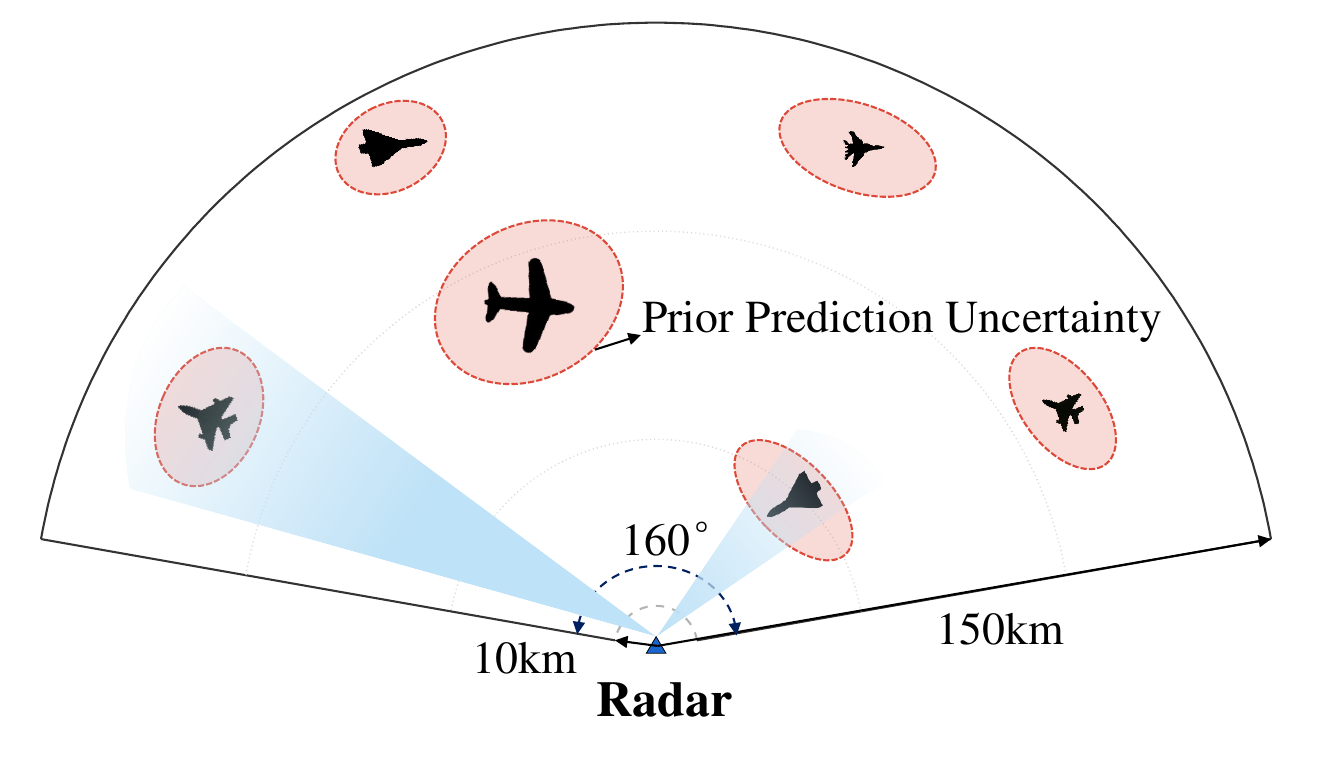}
    \caption{Radar multi-target tracking scenario.}
    \label{fig:environment}
\end{figure}

\noindent \textbf{Environment Setting:} Fig. \ref{fig:environment} illustrates the simulation environment and all targets within the radar's field of view follow the constant-velocity (CV) model. The detailed simulation parameters and symbol definitions are summarized in Table \ref{tab:radar_parameters}.

To verify the generalization capability of the evolved scoring function and prevent data leakage, the scenario datasets used for the AlphaEvolve search in Section \ref{alphaevolve_search} and the performance evaluation in this section are strictly isolated by using independent pseudo-random seeds, while their simulation parameters and corresponding distributions remain identical to those specified in Table \ref{tab:radar_parameters}.

\begin{table}[ht]
    \centering
    \caption{Simulation Parameters and Symbol Definitions}
    \label{tab:radar_parameters}
    \renewcommand{\arraystretch}{1.2} 
    \footnotesize
    \begin{tabular}{lll}
        \toprule
        \textbf{Parameter} & \textbf{Symbol} & \textbf{Value / Distribution} \\
        \midrule
        \multicolumn{3}{l}{\textit{Target-Specific Parameters (for the $i$-th target)}} \\
        Target range & $R_i$ & $\mathcal{U}(10, 150)$ km \\
        Azimuth & $\theta_i$ & $\mathcal{U}(10^\circ, 170^\circ)$ \\
        RCS & $\sigma_i$ & $\mathcal{U}(0.5, 10)$ m$^2$ \\
        Priority weight & $w_i$ & $\mathcal{U}(1, 10)$ \\
        Prior position prediction std. dev. & $\sigma_{p,i}$ & $\text{Log-}\mathcal{U}(10, 10^4)$ m \\
        Prior velocity prediction std. dev. & $\sigma_{v,i}$ & $\text{Log-}\mathcal{U}(1, 100)$ m/s \\
        \midrule
        \multicolumn{3}{l}{\textit{Radar System Parameters}} \\
        Radar scan period & $T_s$ & $1$ s \\
        Total transmit power & $P$ & $2 \times 10^6$ W \\
        Transmit/receive antenna gain & $G$ & $1000$ \\
        Operating wavelength & $\lambda$ & $1$ m \\
        Antenna size constant & $D$ & $5$ m \\
        Transmit baseband bandwidth & $B$ & $10^6$ Hz \\
        Receiver noise bandwidth & $B_n$ & $1.1 \times 10^6$ Hz \\
        Nominal temperature & $T_0$ & $290$ K \\
        Receiver noise figure & $F$ & $10^{0.2}$ \\
        \midrule
        \multicolumn{3}{l}{\textit{Physical Constants}} \\
        Boltzmann constant & $k$ & $1.38 \times 10^{-23}$ J/K \\
        Speed of light & $c$ & $3 \times 10^8$ m/s \\
        \bottomrule
        \multicolumn{3}{p{0.95\columnwidth}}{\scriptsize \textit{Note:} $\mathcal{U}(a, b)$ and $\text{Log-}\mathcal{U}(a, b)$ denote the uniform and log-uniform distributions over $[a, b]$.} \\
    \end{tabular}
\end{table}

\subsection{Approximation Accuracy Analysis}
In this subsection, the solution of the optimization problem obtained by the IPM  (implemented via MATLAB's fmincon solver) is used as the theoretical optimal benchmark (i.e., 0\% error ceiling). Three iterative-free fast power allocation strategies are horizontally compared: uniform allocation, high SNR approximation, and the discovered formula in this paper. Specifically, the high SNR approximation assumes that the current measurement information strictly dominates the posterior information, allowing the prior information matrix $\mathbf{J}_{prior,i}$ to be neglected. Under this assumption, the closed-form power allocation for the $i$-th target is analytically given by 
\begin{equation}
p_i \approx P{\left( w_i \sqrt{\mathrm{Tr}_p\!\left(\widetilde{\mathbf{J}}_{d,i}^{-1}\right)} \right)^{\frac{2}{3}}}/{\sum_{j=1}^N \left( w_j \sqrt{\mathrm{Tr}_p\!\left(\widetilde{\mathbf{J}}_{d,j}^{-1}\right)} \right)^{\frac{2}{3}}}
\label{eq:high snr}
\end{equation}

A total of $10^4$ independent simulation scenarios are generated with the number of targets $N$ uniformly sampled within the range $[10,30]$.
The relative performance loss defined in \eqref{eq:delta_L} is used as the evaluation metric. 

\begin{figure}[htbp]
    \centering
\includegraphics[width=0.45\textwidth]{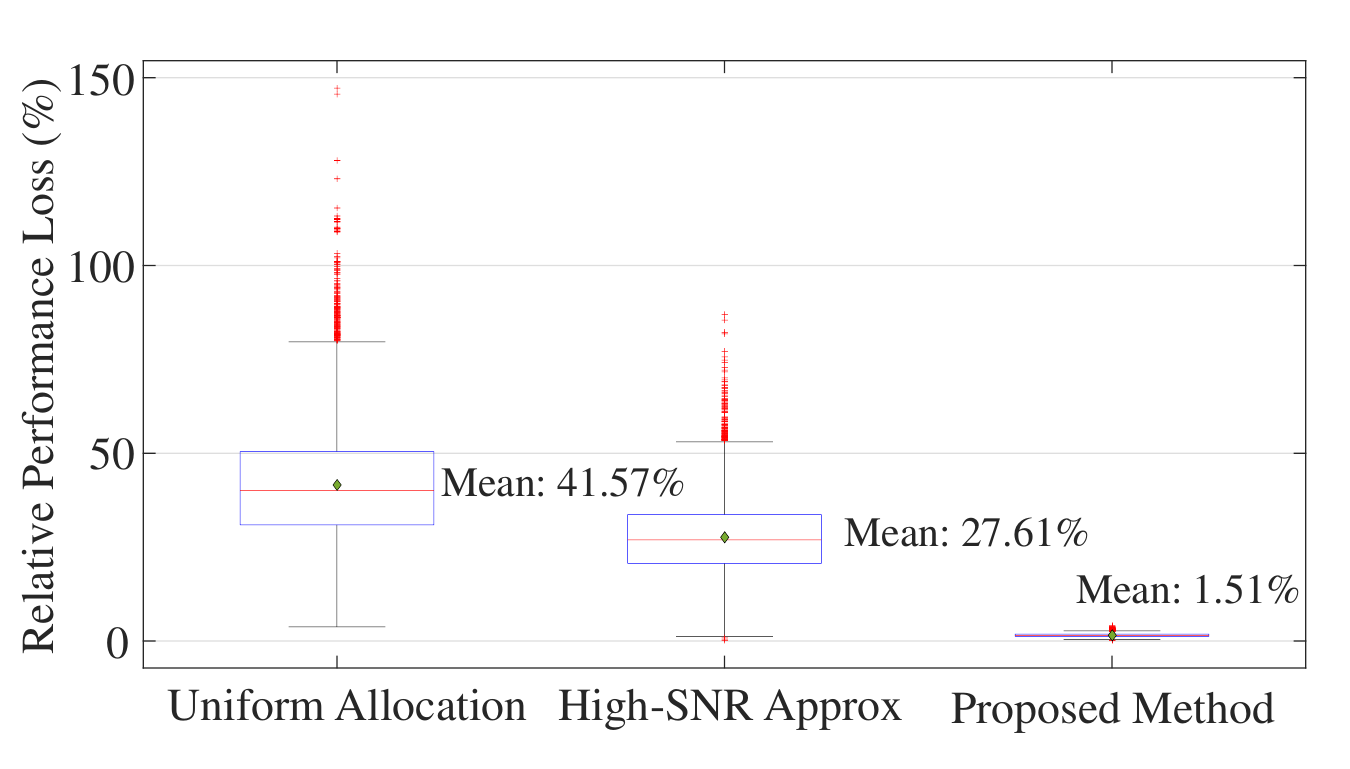}
    \caption{Absolute accuracy and long-tail robustness.}
    \label{fig:xiangxingtu}
\end{figure}

As shown in Fig. \ref{fig:xiangxingtu}, the uniform allocation method completely ignores target differences, resulting in an average relative performance loss as high as $41.57\%$ with a large number of outliers. Although the high SNR approximation method introduces target state information, its accuracy is still insufficient, with an average relative performance loss of $27.61\%$ and a large number of outliers. In contrast, the discovered formula in this paper, under completely iterative-free conditions, has an average relative performance loss of only $1.51\%$ . More importantly, the highly compact distribution of the box plot reveals that even in the most challenging random scenarios, the discovered formula does not suffer from severe performance degradation, demonstrating its exceptional capability to approximate the optimal solution.

\subsection{Ultra-Large-Scale Scenario Generalization Ability}

A core criterion for evaluating the effectiveness of the closed-form scoring function~\eqref{eq:Si} is whether numerical overfitting occurs. Thus we test the closed-form scoring function~\eqref{eq:Si} evolved based on specific scenarios in crowded scenarios with the number of targets $N=10-200$, taking a test point every 5 target numbers, and $10^4$ independent scenarios for each test point.
\begin{figure}[htbp]
    \centering
\includegraphics[width=0.45\textwidth]{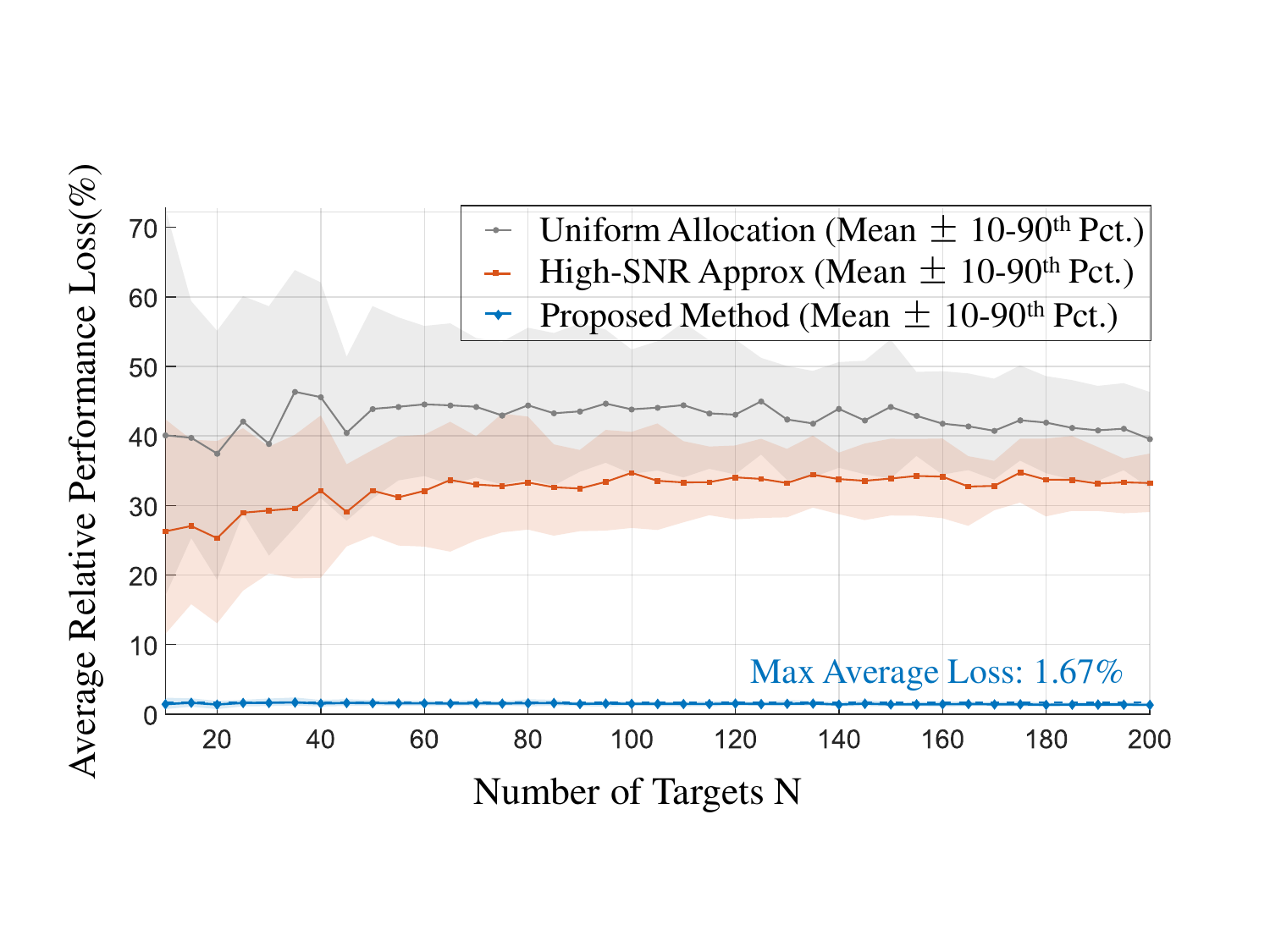}
    \caption{Generalization under scale expansion.}
    \label{fig:fanhua}
\end{figure}

As shown in Fig. \ref{fig:fanhua}, the semi-transparent shadow band (10th-90th Percentile) in the figure excludes $20\%$ of accidental outlier scenarios, intuitively showing the true performance fluctuation range of the discovered formula in $80\%$ of normal scenarios, and the solid line represents the average relative performance loss. With the increase in the number of targets, the error shadow bands of uniform allocation and high SNR approximation diverge sharply; while the maximum average relative performance loss of the discovered formula in this paper is firmly locked at $1.67\%$ throughout the entire scale expansion range, and the error shadow band is also notably narrow. This demonstrates that the discovered formula in this paper has strong scenario generalization and scalability.

\subsection{Computational Efficiency Comparison}
\begin{figure}[htbp]
    \centering
\includegraphics[width=0.45\textwidth]{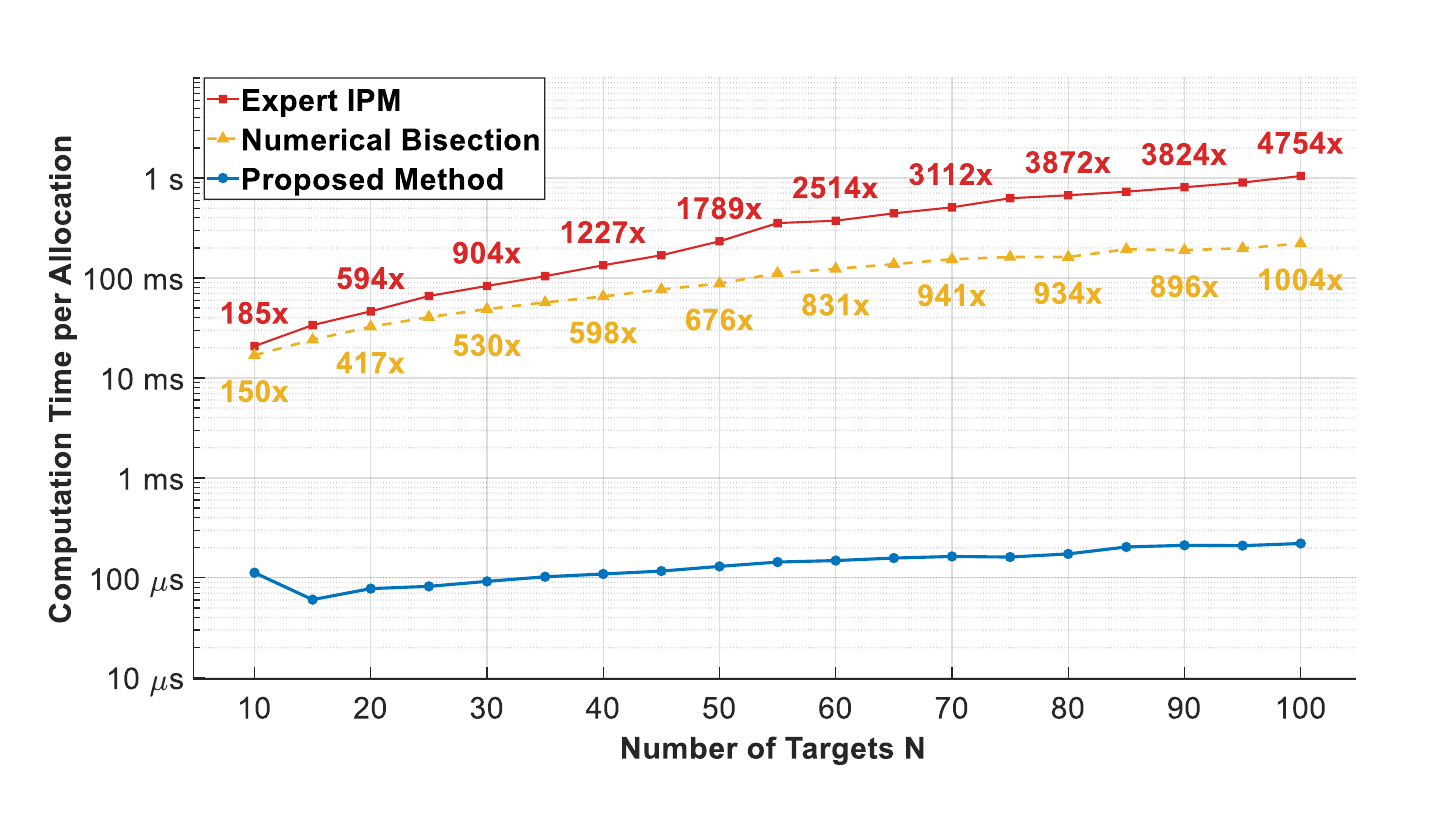}
    \caption{Computation time and speedup ratio.}
    \label{fig:time}
\end{figure}

Computational complexity is a bottleneck restricting real-time scheduling of radar resources. Fig. \ref{fig:time} shows the variation trend of single power allocation time with the number of targets $N$ for different methods. This test is conducted in scenarios with the number of targets $N=10-100$, taking a test point every 5 target numbers, $10^4$ random scenarios for each test point, and running $1000$ times per scenario to take the average time consumption.

As shown in Fig.~\ref{fig:time}, as the number of targets increases, the computation time of both the IPM and the numerical bisection method grows significantly. Furthermore, the speedup ratio of the discovered formula over these two algorithms also increases substantially. In the scenario with $N=100$, the time consumption of the IPM is greater than$1\text{ }s$, and that of the numerical bisection method is greater than $200\text{ }ms$. In contrast, the operation mode of the discovered formula in this paper makes its speed reach $1004$ times that of the numerical bisection method and $4754$ times that of the IPM. In addition, only the time consumption curve of the discovered formula is always below $300~\mu\mathrm{s}$, realizing engineering-level real-time power allocation.

\subsection{Dynamic Closed-Loop Tracking Performance}

\begin{figure}[htbp]
    \centering
\includegraphics[width=0.45\textwidth]{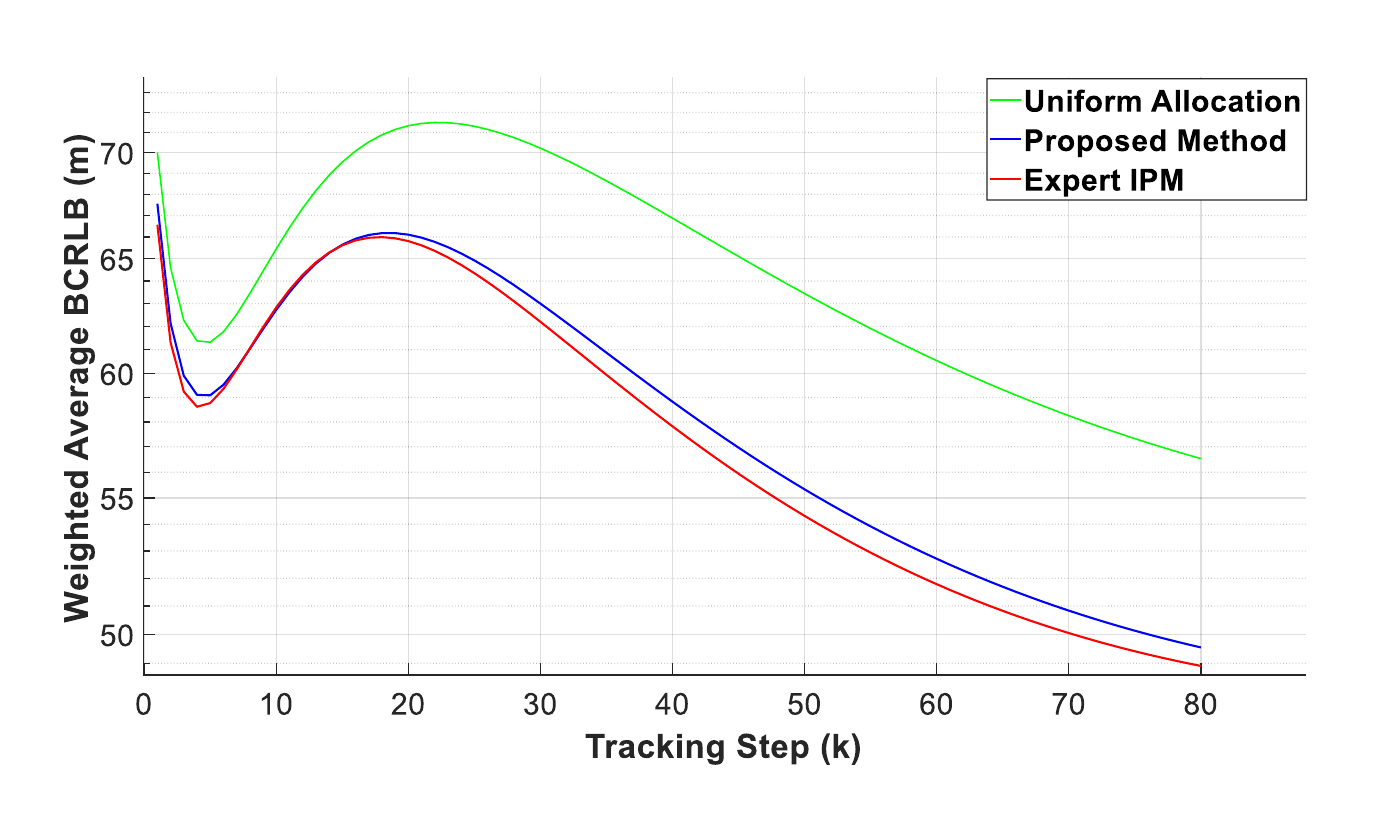}
    \caption{Dynamic BCRLB.}
    \label{fig:BCRLB}
\end{figure}

\begin{figure}[htbp]
    \centering
\includegraphics[width=0.45\textwidth]{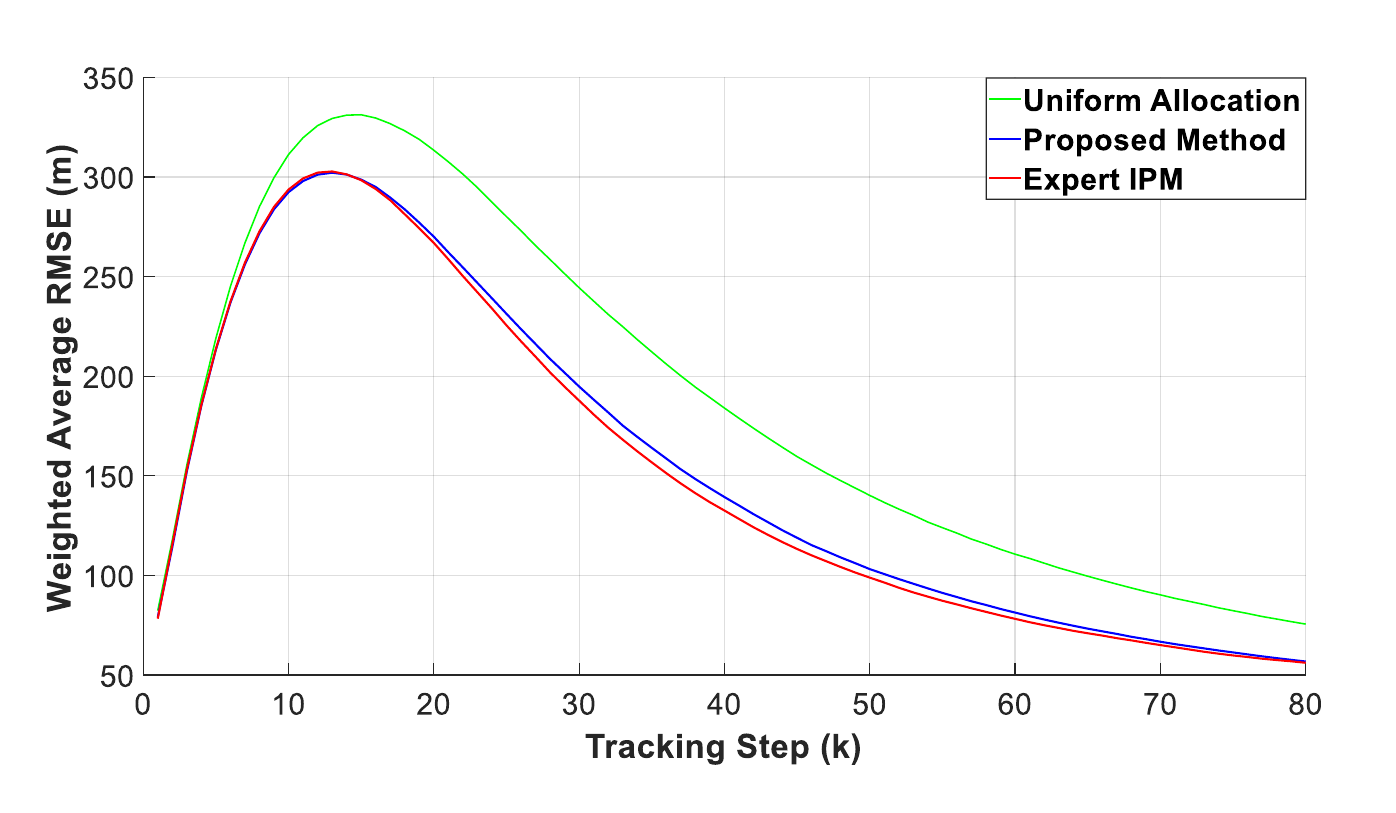}
    \caption{Dynamic RMSE.}
    \label{fig:RMSE}
\end{figure}

To verify the cumulative impact of single-frame power allocation errors in full-life-cycle tracking, Fig. \ref{fig:BCRLB} and \ref{fig:RMSE} present the dynamic EKF tracking performance for up to 80 steps when the number of targets $N=30$, with a total of $10^4$ Monte Carlo cycles.

It can be seen from Fig. \ref{fig:BCRLB} and \ref{fig:RMSE} that whether on the weighted average BCRLB curve evaluating the theoretical lower bound or the weighted average root mean square error (RMSE) curve evaluating the actual filtering accuracy, the discovered formula in this paper has small differences from the IPM throughout the entire time series trajectory. This convincingly demonstrates that the $1.51\%$ average relative performance loss in static optimization does not induce filter divergence in the long-term dynamic closed-loop system, thereby achieving an optimal balance between computational efficiency and near-optimal accuracy.

\section{Conclusion}

This work addresses the computational bottleneck of iterative optimization in radar power allocation by leveraging the search capabilities of large language models to discover a closed-form, approximate power allocation solution. The discovered formula achieves near-optimal performance, with an average relative performance loss of only $1.51\%$ compared to the optimal solution, while delivering a speedup of more than three orders of magnitude. Extensive experiments demonstrate that the discovered formula achieves robust generalization across a wide range of scenarios and target counts, all without the reliance on massive training datasets. These results highlight the potential of LLM-guided symbolic search to revolutionize not only radar resource management but also broader classes of engineering optimization problems.

% This paper has established iteration-free power allocation paradigm for single-radar multi-target tracking. We formulated the resource scheduling as a symbolic search task driven by the AlphaEvolve framework. Our findings confirm that by encoding high-dimensional radar states into physically inspired features, evolutionary program search can autonomously discover a highly compact, closed-form scoring function which can be transformed into the power allocation strategy. Experimental results demonstrate that the discovered formula delivers near-optimal tracking precision and robust generalization while operating at exceptional computational speeds. Furthermore, it reliably avoids filter divergence during prolonged dynamic closed-loop operations.

%This study still has the following limitations: First, it is only carried out for single-radar multi-target power allocation scenarios, and does not cover more complex multi-dimensional resource management problems. Second, feature extraction still relies on manual domain prior knowledge and the automatic embedding ability of physical prior knowledge need to be further improved.

%Future work will further extend this paradigm to more complex resource management scenarios, realize the automatic embedding of physical prior knowledge to reduce manual prior dependence and combine the domain knowledge reasoning ability of large language models to optimize symbolic search efficiency.

\appendix  % 切换为附录模式

\section{DERIVATIONS OF THE BAYESIAN INFORMATION MATRIX COMPONENTS}
\label{app:BIM}

In this appendix, we detail the explicit formulations of the normalized measurement information matrix $\widetilde{\mathbf{J}}_{d,i}$ (i.e., the information obtained per unit transmit power) and the prior information matrix $\mathbf{J}_{\text{prior},i}$ used in the computation of the BCRLB.

For the $i$-th target, the radar measurement model relates the target state $\mathbf{x}_i = (x_i, \dot{x}_i, y_i, \dot{y}_i)$ to the range and azimuth measurements. The nonlinear measurement function $h(\mathbf{x}_i)$ is
\begin{equation}
h(\mathbf{x}_i) = \begin{bmatrix} r_i \\ \theta_i \end{bmatrix} 
= \begin{bmatrix} \sqrt{x_i^2 + y_i^2} \\[4pt] \arctan\!\left(\dfrac{y_i}{x_i}\right) \end{bmatrix}
\end{equation}
where $r_i$ and $\theta_i$ denote the true range and azimuth, respectively.

Let $\hat{\mathbf{x}}_{i,k|k-1}$ be the predicted state of target $i$ at time step $k$. The Jacobian matrix of $h$ with respect to $\mathbf{x}_i$, evaluated at this predicted state, is required to construct the measurement information matrix. Taking the partial derivatives yields the $2 \times 4$ Jacobian
\begin{equation}
\mathbf{H}_i = \left.\frac{\partial h(\mathbf{x}_i)}{\partial \mathbf{x}_i}\right|_{\mathbf{x}_i = \hat{\mathbf{x}}_{i,k|k-1}}
= \begin{bmatrix}
\frac{x_i}{r_i} & 0 & \frac{y_i}{r_i} & 0 \\[4pt]
-\frac{y_i}{r_i^2} & 0 & \frac{x_i}{r_i^2} & 0
\end{bmatrix}_{\mathbf{x}_i = \hat{\mathbf{x}}_{i,k|k-1}} 
\end{equation}

Following the radar equation and the power allocation model in the main text, the measurement covariance matrix for target $i$ can be expressed as a function of the allocated power $p_i$ :
\begin{equation}
\boldsymbol{\Sigma}^i(p_i) = \mathrm{diag}\bigl(\sigma_{r,i}^2,\; \sigma_{\theta,i}^2\bigr)
= \mathrm{diag}\!\left(\frac{\gamma_{r,i}}{p_i},\; \frac{\gamma_{\theta,i}}{p_i}\right)
= \frac{1}{p_i}\,\widetilde{\boldsymbol{\Sigma}}^i
\label{eq:cov_factor}
\end{equation}
where the power-normalized covariance matrix $\widetilde{\boldsymbol{\Sigma}}^i$ is defined as
\begin{equation}
\widetilde{\boldsymbol{\Sigma}}^i = \mathrm{diag}(\gamma_{r,i},\; \gamma_{\theta,i})
\end{equation}
The terms $\gamma_{r,i}$ and $\gamma_{\theta,i}$ encapsulate target-specific and system-specific parameters (e.g., RCS and antenna gain), which are independent of the dynamically allocated power $p_i$.

Consequently, the unit-power normalized measurement information matrix, denoted by $\widetilde{\mathbf{J}}_{d,i}$, is given by the standard Fisher information formula applied to the normalized covariance:
\begin{equation}
\widetilde{\mathbf{J}}_{d,i} = \mathbf{H}_i^{\mathrm{T}} \bigl(\widetilde{\boldsymbol{\Sigma}}^i\bigr)^{-1} \mathbf{H}_i
\label{eq:Jd_tilde}
\end{equation}
Physically, $\widetilde{\mathbf{J}}_{d,i}$ represents the measurement information that would be obtained if a unit of transmit power were allocated to target $i$.

Substituting the expressions for $\mathbf{H}_i$ and $(\widetilde{\boldsymbol{\Sigma}}^i)^{-1} = \mathrm{diag}(\gamma_{r,i}^{-1},\, \gamma_{\theta,i}^{-1})$ into~\eqref{eq:Jd_tilde} and carrying out the matrix multiplication, we obtain the explicit $4 \times 4$ form
\begin{equation}
\widetilde{\mathbf{J}}_{d,i} = 
\begin{bmatrix}
\Xi_{11}^i & 0 & \Xi_{13}^i & 0 \\
0 & 0 & 0 & 0 \\
\Xi_{31}^i & 0 & \Xi_{33}^i & 0 \\
0 & 0 & 0 & 0
\end{bmatrix}
\label{eq:Jtilde_explicit}
\end{equation}
where the non-zero entries are
\begin{align}
\Xi_{11}^i &= \frac{x_i^2}{\gamma_{r,i} r_i^2} + \frac{y_i^2}{\gamma_{\theta,i} r_i^4} \label{eq:Xi11} \\
\Xi_{33}^i &= \frac{y_i^2}{\gamma_{r,i} r_i^2} + \frac{x_i^2}{\gamma_{\theta,i} r_i^4} \label{eq:Xi33} \\
\Xi_{13}^i &= \Xi_{31}^i = \frac{x_i y_i}{\gamma_{r,i} r_i^2} - \frac{x_i y_i}{\gamma_{\theta,i} r_i^4} \label{eq:Xi13}
\end{align}
All state variables in~\eqref{eq:Xi11}--\eqref{eq:Xi13} are taken at the predicted state $\hat{\mathbf{x}}_{i,k|k-1}$. The zero rows and columns reflect the fact that the range and azimuth measurements do not directly observe the velocity components $(\dot{x}_i,\dot{y}_i)$.

$\mathbf{J}_{\text{prior},i}$ encapsulates the knowledge available before the current measurement is incorporated. In a standard tracking framework such as the Extended Kalman Filter (EKF), it is given by the inverse of the predicted state covariance matrix at time $k$:
\begin{equation}
\mathbf{J}_{\text{prior},i} = \bigl(\mathbf{P}_{k|k-1}^i\bigr)^{-1}
\end{equation}
where $\mathbf{P}_{k|k-1}^i$ is the error covariance matrix of the predicted state $\hat{\mathbf{x}}_{i,k|k-1}$, obtained from the kinematic motion model and the previous filtering step.

Finally, adding the prior information and the measurement information scaled by the allocated power $p_i$ yields the complete posterior Bayesian information matrix of target $i$, as stated in the main text:
\begin{equation}
\mathbf{J}_i(p_i) = \mathbf{J}_{\text{prior},i} + p_i \cdot \widetilde{\mathbf{J}}_{d,i}.
\end{equation}

\section{KKT CONDITIONS AND PHYSICALLY INSPIRED FEATURES DERIVATION}
\label{app:kkt_features}
In this appendix, we theoretically justify the extraction of the physically inspired features, specifically the demand factor $D_i$ and the marginal benefits $M_{base,i}$ and $M_{cliff,i}$, by analyzing the KKT conditions of the power allocation problem.

The optimization problem $\mathcal{P}$ seeks to minimize the total weighted BCRLB under the total power constraint $\sum p_i = P$ and the minimum power constraints $p_i \ge p_{min}$. The Lagrangian function is constructed as:
\begin{equation}
\begin{split}
\mathcal{L}(\mathbf{p}, \lambda, \boldsymbol{\mu}) = & \sum_{i=1}^N w_i \cdot \mathrm{BCRLB}_i(p_i) + \lambda \left( \sum_{i=1}^N p_i - P \right) \\
& + \sum_{i=1}^N \mu_i (p_{min} - p_i)
\end{split}
\end{equation}
where $\lambda$ and $\mu_i \ge 0$ are the Lagrange multipliers. The KKT stationarity condition requires that at the optimal solution $p_i^*$:
\begin{equation}
\frac{\partial \mathcal{L}}{\partial p_i} = w_i \frac{\partial \mathrm{BCRLB}_i(p_i)}{\partial p_i} + \lambda - \mu_i = 0
\label{eq:kkt_stationarity}
\end{equation}

According to the KKT complementary slackness condition $\mu_i (p_{min} - p_i) = 0$, if a target is allocated an optimal power strictly exceeding the minimum threshold (i.e., $p_i^* > p_{min}$), the constraint becomes inactive, which mathematically enforces the corresponding multiplier to be zero ($\mu_i = 0$). In practical radar resource management, this inactive constraint condition naturally holds when the system operates with an abundant total power budget, or for primary targets with high tracking demands (e.g., high priority weights or challenging observation geometries) that require substantial power illumination.

Simultaneously, such sufficient power allocation places these targets into a \textit{high-SNR regime}. In this regime, the real-time measurement information strictly dominates the prior prediction information, allowing the posterior information matrix to be simplified as $\mathbf{J}_i(p_i) \approx p_i \tilde{\mathbf{J}}_{d,i}$. Under these concurrent conditions, the target BCRLB can be analytically approximated as:
\begin{equation}
\mathrm{BCRLB}_i(p_i) \approx p_i^{-1/2} \sqrt{ \mathrm{Tr}_p \! \left( \tilde{\mathbf{J}}_{d,i}^{-1} \right) }
\end{equation}
Substituting this approximated objective and the condition $\mu_i = 0$ into the stationarity equation \eqref{eq:kkt_stationarity}, we obtain:
\begin{equation}
-\frac{1}{2} p_i^{-3/2} w_i \sqrt{ \mathrm{Tr}_p \! \left( \tilde{\mathbf{J}}_{d,i}^{-1} \right) } + \lambda = 0
\end{equation}
Solving for $p_i$, the allocation strategy is analytically given by:
\begin{equation}
p_i \propto \left( w_i \sqrt{ \mathrm{Tr}_p \! \left( \tilde{\mathbf{J}}_{d,i}^{-1} \right) } \right)^{2/3}
\end{equation}
This rigorous derivation explicitly demonstrates that the $2/3$ exponent is an intrinsic mathematical property of the radar covariance inversion. However, the aforementioned high-SNR approximation entirely omits the prior prediction information. To compensate for this omission, we reintroduce the prior information by evaluating the complete posterior information matrix at an equal-sharing anchor point $\bar{p} = P/N$. This physically grounded adjustment yields the exact formulation of the absolute demand factor $D_i$ defined in \eqref{eq:Di}, which can also serves as a theoretically optimal seed to narrow the immense search space of the AlphaEvolve framework.

Furthermore, for generalized scenarios where the high-SNR approximation does not strictly hold, the term $w_i \frac{\partial \mathrm{BCRLB}_i(p_i)}{\partial p_i}$ in \eqref{eq:kkt_stationarity} represents the true descent gradient of the objective function. By evaluating the absolute value of this gradient at the equal-sharing point $\bar{p}$ and the minimum threshold $p_{min}$, we precisely obtain the baseline marginal benefit $M_{base,i}$ and the cliff marginal benefit $M_{cliff,i}$ defined in ~\eqref{eq:MR_base}--\eqref{eq:MR_cliff}.

\end{document}